\documentclass[twocolumn,showpacs,amsmath,amssymb]{revtex4}

\usepackage{graphicx}
\usepackage{dcolumn}
\usepackage{bm,color}

\begin{document}

\preprint{APS/}

\title{Energy-dependent existence of soliton in the synthesis of chemical elements}

\author{Yoritaka Iwata$^{1}$}
 \email{iwata@cns.s.u-tokyo.ac.jp}

 \affiliation{$^{1}$School of Science, The University of Tokyo, 113-0033 Tokyo, Japan
}

\date{\today}

\begin{abstract}
Light chemical elements are, for instance, produced through ion collisions taking place in the core of stars, where fusion is particularly important to the synthesis of chemical elements.
Meanwhile soliton provides non-interacting transparency leading to the hindrance of fusion cross section.
In order to explain high fusion cross section actually observed in low incident energies, it is necessary to discover the suppression mechanism of soliton propagation.
In this paper, based on a systematic three-dimensional time-dependent
density functional calculation, the existence of soliton is examined for ion
collisions with some incident energies, impact parameters, and nuclear force
parameter sets.
As a result solitons are suggested to exist highly depending on the energy. 
The suppression of soliton is consequently due to the spin-orbit force and the momentum-dependent components of the nuclear force.
\end{abstract}

\maketitle

\section{Introduction}
Nuclei are fundamental pieces of building up galaxies, the earth, and all
the creatures and materials around us.
A nucleus consists of neutrons and protons, and the specie of chemical element is identified by the number of protons contained in a nucleus, where the same chemical elements with the different neutron numbers are called isotopes.
The synthesis of certain chemical elements from the other chemical elements has not been industrially realized in
factories so far, but naturally taking place in the universe (e.g., in the core of stars).
In those sites low-energy (heavy-)ion reactions, which includes several types of reactions such as fusion, fragmentation, fission, and inelastic scattering, give rise to the synthesis of chemical elements.
It is worth noting that the typical incident energy of producing
chemical elements is as much as a few MeV per nucleon, where the neutron and the
proton are collectively called the nucleon.

Soliton is a concept arising from the nonlinearity and the dispersive
property  (for the general argument about soliton, see textbooks such as
\cite{11ablowitz}), where the dispersive property is represented by
$\omega \ne c_v k$ for the angular frequency $\omega$, the wave number $k$,
and a constant $c_v$.
In the literature of soliton research (for a milestone, see \cite{65zabusky}), the stably propagating wavepacket is called solitary wave, and such a solitary
wave is called soliton if its essential properties such as mass and momentum distributions are conserved throughout the time evolution (even after the collisions).
A wide appearance of soliton regardless of the size of the system has been
known, which suggests the underlying mathematically universal property.
While soliton were historically studied in many complex systems such
as shallow water wave described by KdV equation \cite{1895kdv}, the preceding soliton studies associated with the ion collisions (including nuclear fusion reaction) are quite limited.
Indeed, although the soliton in heavy-ion reactions was studied in
Raha-Weiner~\cite{83raha} with paying attention to the momentum transparency, the suppression mechanism of soliton in ion collisions has been an open problem.
In particular theoretical research taking into account realistic conditions like three-dimensional motion is claimed to be desired in Ref.~\cite{83raha}.
Note that the theoretical soliton studies on the sub-atomic
physics are rather actively developed in the field of the particle physics and
the hadron physics, but those energies are too high to have something to do
with the synthesis of chemical elements.

In this paper, based on the three-dimensional density functional theory (TDDFT), the appearance of soliton in sub-atomic reaction process on the scale of 10$^{-14}$~m is studied.
The appearance of soliton suggests the actual influence of a certain kind of nonlinearity
in the synthesis of chemical elements, while the appearance of fusion is exactly the contour example for the stable existence of soliton. 
Consequently, the answer to the question ''Can nuclei (many-nucleon systems) be solitons ?" is presented.

\section{Numerical experiment}
\subsection{Ion reaction producing light chemical elements}
For the synthesis of chemical elements, we take collisions between two
helium isotopes:
\begin{equation} \label{reac}
^4{\rm He} + ^8{\rm He} 
\end{equation}
where $^4$He consists of 2 protons and 2 neutrons, and $^8$He consists of 2
protons and 6 neutrons.
Although there should be a neutron-rich situation for the existence of $^8$He, this reaction can take place in helium-abundant stars.

Even restricted to the collision shown in Eq.~\eqref{reac}, there should be several
types of reactions such as fusion, fragment, fission, and inelastic scattering.
A wide variety of reactions are realized by setting different initial conditions: the incident energy and the impact parameter.
If fusion appears, the final product is $^{12}$Be including 4 protons and 8 neutrons.
If reactions other than fusion appears, fragments smaller than $^{12}$Be can be produced.
That is, hydrogen (proton number=$1$), helium (proton number=$2$), lithium (proton number=$3$), and
beryllium (proton number=$4$) isotopes are possible to be produced through this reaction.

\subsection{Theoretical model}
One of the most reliable frameworks of describing low-energy (heavy-)ion
collisions is the nuclear time-dependent density functional theory \cite{30dirac,75engel}.
In case of Eq.~\eqref{reac}, the non-stationary problem is written by
\begin{equation}  \label{nonstationaryaryeq}  
i \hbar \frac{\partial}{\partial t} \psi_{q,j}(r,t) = H(\psi_{q,j}(r,t)) ~
\psi_{q,j}(r,t), \quad j=1,\cdots,12,  
\end{equation}
where $\psi_{q,j}(r,t)$ denotes single nucleon wave function ($r \in {\mathbb R}^3$, $t \in {\mathbb R}$), the index $q$ identifies protons ($q=$-1/2) and neutrons ($q=$+1/2),  and $H(\psi_{q,j}(r,t)) $ is the effective Hamiltonian operator
\begin{equation} \label{hamilton} \begin{array}{ll}
 H(\psi_{q,j}(r,t)) = K + V_{q}(\psi_{q,j}(r,t)),  \vspace{2.5mm}\\
 V_q(\psi_{q,j}(r,t)) =
V_{N,q}(\psi_{q,j}(r,t)) + V_{C,q}(\psi_{q,j}(r,t)), 
\end{array} \end{equation}
which consists of the kinetic part $K$ and the potential part $V_{q}$. 
Equation~\eqref{nonstationaryaryeq} is a nonlinear equation holding the
dispersive property ($\omega \ne c_v k$) in which the single-nucleon degrees
of freedom are fully taken into account.
Spin degree of freedom is also taken into account, although it is not explicitly shown.
The stationary problem is written by
\begin{equation} \label{stationaryeq}  H(\psi_{q,j}(r,t)) \psi_{q,j}(r,t)  = 0,  \end{equation}
where the condition $\partial \psi_{q,j}(r,t)/\partial t = 0$ is additionally imposed to Eq.~\eqref{nonstationaryaryeq}.

\begin{table}
 \caption{Coefficient values of the SV-bas interaction parameter set \cite{07klupfel}.
This parameter set is recently introduced in terms of obtaining a good
description of the giant resonances in $^{208}$Pb. 
\vspace{1.5mm}}
  \begin{tabular}{|c||c|} \hline
   Parameter &  Value \\   \hline 
   $t_0$(MeV$\cdot$fm$^3$) & -1879.640018  \\   \hline
   $t_1$(MeV$\cdot$fm$^5$)  & 313.7493427  \\   \hline
   $t_2$(MeV$\cdot$fm$^5$)  & 112.6762700    \\   \hline
   $t_3$(MeV$\cdot$fm$^{3+3\alpha}$)  & 12527.38921   \\   \hline
   $W_0$(MeV$\cdot$fm$^5$)  & 124.6333000    \\   \hline    
     \end{tabular}  
       \begin{tabular}{|c||c|} \hline
       Parameter &  Value \\   \hline 
   $x_0$ & 0.2585452462    \\   \hline
   $x_1$ & -0.3816889952    \\   \hline
   $x_2$ & -2.823640993   \\   \hline 
   $x_3$ & 0.1232283530  \\   \hline
   $\alpha$ & 0.30  \\ \hline
  \end{tabular}  \vspace{1.5mm} \\
 \label{table0}
\end{table}

\begin{table}
 \caption{The stationary state energy (MeV) for $^4$He and $^8$He using different interaction sets.
There is no bound states for the interactions with $\lambda_0 = 0$, because those energies cannot be negative.
Some energies shown in parentheses mean that the iteration is not convergent at the order of 0.01~MeV.
For reference, the experimental bound state energy for $^4$He and $^8$He are -28.30~MeV and -31.40~MeV, respectively.
\vspace{1.5mm}}
  \begin{tabular}{|c||c|c|c|c||} \hline 
   Stationary state energy & $^4$He &  $^8$He \\   \hline 
   SVb10000 & -1103.87  & -3160.90   \\   \hline
   SVb10001 & -1103.51  & (-4601.36)   \\   \hline
   SVb10010 & -43.30  & -49.25   \\   \hline
   SVb10011 & -43.27  & -61.75   \\   \hline
   SVb10100 & -1700.32  & (-5034.96)   \\   \hline
   SVb10101 & -1700.32  & (-5631.16)   \\   \hline
   SVb10110 & -43.49  & -824.92   \\   \hline
   SVb10111 & -43.44  & -1848.16   \\   \hline
   SVb11000 & -595.00 & -1402.98   \\   \hline
   SVb11001 & -594.09  & -1917.26   \\   \hline
   SVb11010 & -16.20  & -15.36   \\   \hline
   SVb11011 & -16.20  & -17.03   \\   \hline
   SVb11100 & -1253.08  & -2242.38   \\   \hline
   SVb11101 & -1252.99  & -2974.18   \\   \hline
   SVb11110 & -16.20  & -18.96   \\   \hline
   SVb11111 & -16.20  & -22.77   \\   \hline
  \end{tabular}  \vspace{1.5mm} \\
 \label{table1}
\end{table}

\begin{figure*} 
 \includegraphics[width=12.0cm]{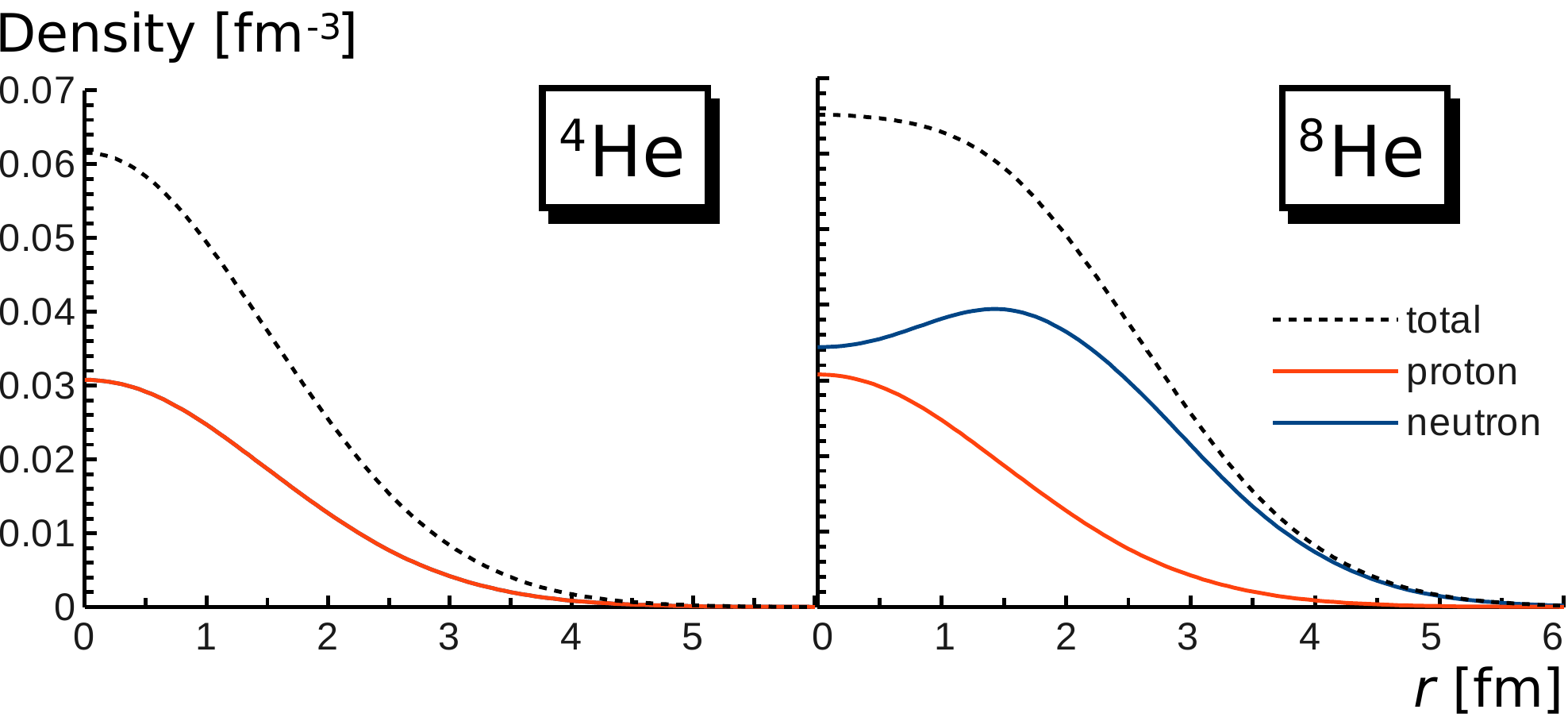}
\caption{(Color online) The radial dependence of the density distribution for the bound states of
 $^4$He and $^8$He employing SVb11111 interaction. 
 The total (dotted line), proton (red line) and neutron (blue line) densities are depicted as a function of the radial distance $r$.
 The obtained discrete data are connected using the spline interpolation.
} 
\label{fig1}
\end{figure*}

Specific features of the density functional theory can be found in the form of unknown function and the form of interaction.
First, the unknown many-body wave function is of the form of the Slater determinant; i.e., Pauli principle is taken into account.
Second, the nuclear potential part $V_{N,q}$ of the effective Hamiltonian is obtained by applying the variational principle to the Skyrme's zero-range interaction \cite{Skyrme}:
\[ \begin{array}{ll}
v_{i,j}({\bm k},  {\bm k}') = \lambda_0 ~ t_0 (1 + x_0  P_{\sigma}) \delta(r_i -r_j)  \vspace{1.5mm}\\
 + \lambda_1 ~ \frac{t_1}{2} (1 + x_1  P_{\sigma}) \{ \delta(r_i -r_j)  {\bm k}^2 +  {\bm k}'^2\delta(r_i -r_j)\}   \vspace{1.5mm}\\
 + \lambda_2 ~  t_2  (1 + x_2  P_{\sigma})  {\bm k}' \delta(r_i -r_j)  {\bm k}  \vspace{1.5mm}\\
 + \lambda_3 ~ \frac{t_3}{6}(1 + x_3 P_{\sigma})
 \sum_j  |\psi_{q,j}((r_i+r_j)/2,t)|^{2 \alpha} \delta(r_i -r_j)   \vspace{1.5mm}\\
 + \lambda_4  ~ i W_0 ({\bar \sigma}_i + {\bar \sigma}_j) \cdot  {\bm k}' \times \delta(r_i -r_j)  {\bm k},
\end{array} \]
where ${\bm k}$ and ${\bm k}'$ are the relative wave vectors of two interacting nucleons,
$t_0$, $t_1$, $t_2$, $t_3$, $x_0$, $x_1$, $x_2$, $x_3$, $\alpha$ and $W_0$ are
parameters (for the actually used values, see Table~\ref{table0}).
Additional parameters $\{ \lambda_i = 0,1; ~ i =0,1,\cdots, 4\}$ are introduced to turn on/off each term.
For instance, in case of $(\lambda_0,\lambda_1,\lambda_2,\lambda_3,\lambda_4)
= (1,0,0,0,0)$, the form of the nuclear potential part is represented by 
\begin{equation} \label{typinhom} \begin{array}{ll}
V_{N,q}(\psi_{q,j}(r,t)) =  \vspace{1.5mm}\\
 t_0 \{ (1+\frac{x_0}{2}) \sum_{j} |\psi_{q,j} (t,r) |^2 
- (\frac{1}{2} + x_0) \sum_{q} |\psi_{q,j} (t,r) |^2 \}  
\end{array} \end{equation}
where $\sum_{j}$ and $\sum_{q}$ denote the sum over all $j$ and that over all
$j$ with a fixed $q$, respectively (for the mathematical derivation of these terms, see
\cite{13iwata-maruhn}).
Terms shown in Eq.~\eqref{typinhom} are the same form as the typical nonlinear Schr\"dinger equations showing the soliton propagation. 
For the functional form of $V_{N,q}(\psi_{q,j}(r,t))$ in case of
$(\lambda_0,\lambda_1,\lambda_2,\lambda_3,\lambda_4) = (1,1,1,1,1)$, see Refs.~\cite{72vautherin, 75engel, 98chabanat, 07klupfel}.
As is readily seen in Eq.~\eqref{typinhom}, the operation of the nuclear force can be different
between neutrons and protons even position $r$ and time $t$ are set to be exactly the
same. 
In particular the term with $W_0$ means the spin-orbit force whose origin can
be found in the special relativity theory.
The terms with $t_0$ and  $t_3$ are independent of the momentum
(${\bm k}$ and ${\bm k}'$).
In addition to the nuclear force, the Coulomb force $V_{C,q}$ is also taken
into account in this theoretical framework (Eq.~\eqref{hamilton}).

Periodic boundary condition is imposed to both stationary and non-stationary
problems, and an initial wave function (at $t=0$) is given to non-stationary problems.
Following a general method used in the theory of low-energy heavy-ion
collisions, the initial momentum distribution is given by multiplying the
plain wave in a certain direction to the initial wave function~\cite{13maruhn} (this process
is sometimes called "boost").
In this manner, traveling waves with certain velocities are prepared as the initial states.

We actually consider 32 different nuclear interactions, where the SV-bas interaction parameter set is taken as the basis interaction.  
Binary numbers are made by lining up $ \lambda_0$, $\lambda_1$,
$\lambda_2$, $\lambda_3$, and $\lambda_4$ in a row.
Those binary numbers are used for labeling 32 different interactions (Table~\ref{table1}); e.g.,
SVb10100 stands for the interaction with $(\lambda_0,\lambda_1,\lambda_2,\lambda_3,\lambda_4) = (1,0,1,0,0)$, and SVb11111 (equal to the original SV-bas interaction parameter set) corresponds to the interaction with
$(\lambda_0,\lambda_1,\lambda_2,\lambda_3,\lambda_4) = (1,1,1,1,1)$.
The Coulomb force is turned-on by default.

\subsection{Numerical setup}
\begin{figure} 
 \includegraphics[width=7.0cm]{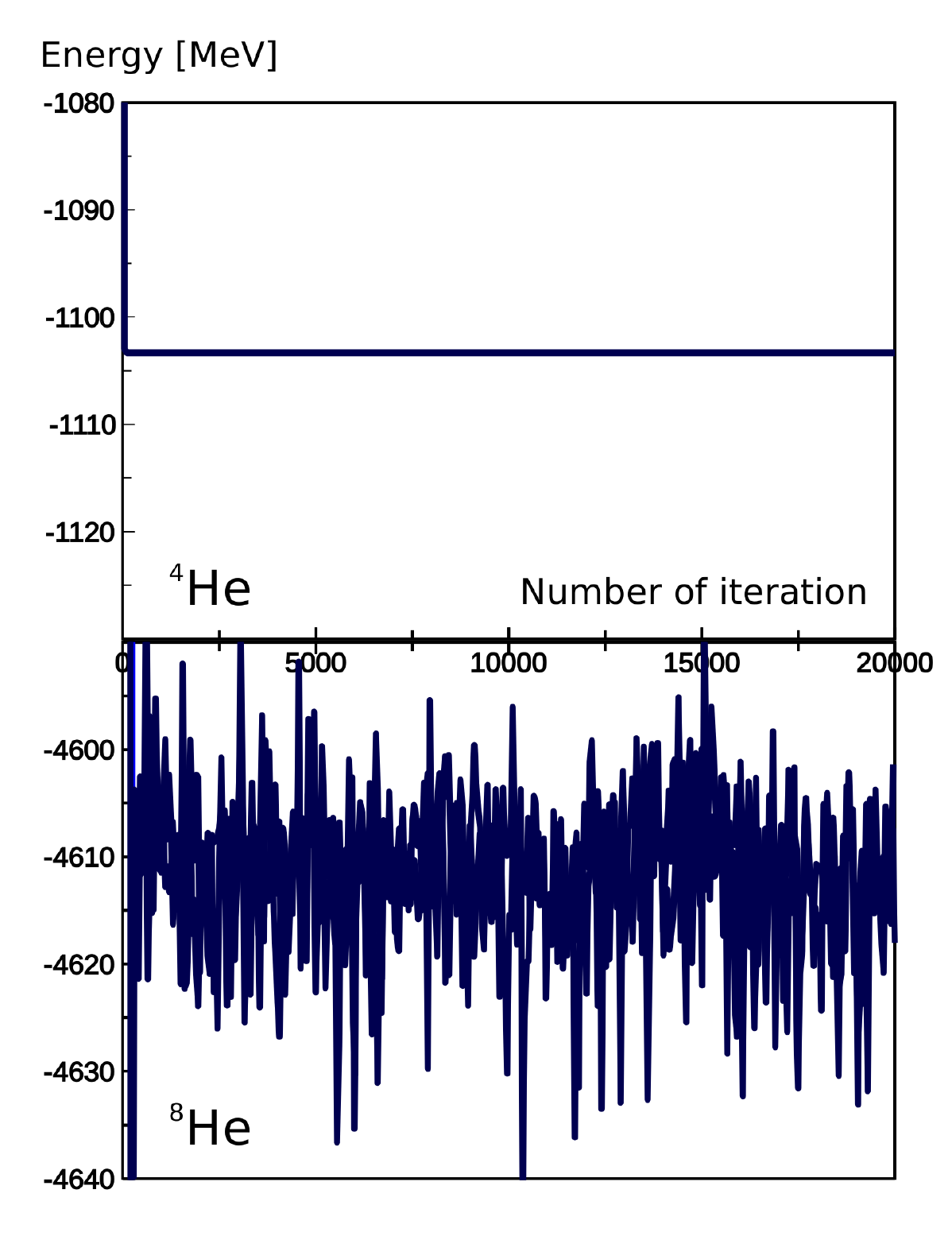}
\caption{(Color online) Convergence of the stationary solution employing SVb10001 interaction.
The bound state is obtained for the upper panel, while it is not for the lower panel.
} 
\label{fig3}
\end{figure}

Numerical solutions are obtained based on the finite difference method (for the details, see ~\cite{13maruhn}). 
Three-dimensional space is incremented by 1.0~fm and the unit time step is set
to one-third of 10$^{-23}$~s.
Vacuum boxes are prepared as 24$\times$ 24$\times$ 24~fm$^3$ for
the stationary problems, and as 32$\times$ 32$\times$ 32~fm$^3$ for the non-stationary problems.
Numerical procedure is explained step by step.
First, the bound states are prepared.
By solving the stationary problem Eq.~\eqref{stationaryeq} with a given interaction parameter set, we obtain the
bound states (self-bound states) of $^4$He and $^8$He (Fig.~\ref{fig1}).
Total 20000 steps of numeral iteration (imaginary time evolution) leading to the convergence are made in
each case, where the convergence is determined if the binding energy (within the limit accuracy of 0.01~MeV) is conserved for 1000 iterations between 19000 and 20000 iterations.
In case of SVb10001 interaction, the stationary solution is not obtained for $^8$He but for $^4$He (Fig.~\ref{fig3}) in which a few hundred iterations lead to the convergence for $^4$He but oscillations with those amplitudes 10~MeV still remain even around 20000 iterations for $^8$He.
A set of calculation has been carried out for 32 different interactions.
Results are summarized in Table~\ref{table1}, where the bound states for $^4$He or $^8$He are not obtained for SVb10001, SVb10100, SVb10101 and the interactions with $\lambda_0 = 0$.
Stationary states are finally determined to be the bound states, if they are stable after changing the size of the inclusive vacuum box.
Consequently, among several interactions including the bound states, SVb11010,
SVb11011, SVb11110, and SVb11111 interaction sets are chosen to go further
into the collision dynamics, because others have unphysically high or low stationary state energies, which are worthless in terms of comparing with the most realistic case employing SVb11111 interaction. 
Indeed, after making non-stationary calculations in a larger box ($36\times36\times36$~fm$^3$), $^8$He for SVb10011 has turned out to be not a bound state but
only a stationary state existing only in a spatial box $24\times24\times24$~fm$^3$.

Second, the initial state is prepared using the bound states.
At $t=0$~s, put $^4$He and $^8$He into the vacuum box; $^4$He is placed at
$(6,b/2,0)$ and $^8$He at $(-6,-b/2,0)$.
Here $b$~fm is an approximation for the impact parameter, and we call
the above $b$ the modified impact parameter.
Three different incident energies are examined; the reduced incident energies
$E_s$ are set to 1.0~MeV, 10.0~MeV and 100.0~MeV, respectively. 
The relative momentum, which is parallel to the $x$-axis, is given to $^4$He
and $^8$He based on the previously-mentioned boost method.
Let the masses of $^4$He and $^8$He be denoted by $M_4$ and $M_8$ respectively,
The initial momentum of $^4$He and $^8$He are generally set to $(-\sqrt{2 M_{4} E_s},0,0)$
and $(\sqrt{2 M_{8} E_s},0,0)$ respectively.
Two different types of collisions are carried out; central collisions ($b=0.0$~fm) and peripheral collisions ($b=4.0$~fm).
By denoting the speed of light by $c$, the amplitude of the initial relative velocities
$v_{rel}$ for the central collisions of $E_s =1.0$~MeV, $E_s = 10.0$~MeV, and
$E_s = 100.0$~MeV cases are 0.09$c$, 0.28$c$, and 0.81$c$ respectively.
The duration time for the cases with three different energies is estimated by
\begin{equation} \begin{array}{ll} \label{timestim}
 t = \frac{(4^{1/3} + 8^{1/3}) r_0} {v_{rel}}, \quad  r_0=1.2~{\rm fm} . 
\end{array} \end{equation} 
They are roughly equal to 2.7$\times10^{-22}$s, 0.7$\times10^{-22}$s, and 0.2$\times10^{-22}$s for $E_s=1.0$~MeV, 10.0~MeV and 100.0~MeV, respectively.
Since the investigation of the transparency is the evaluation of interaction between the two bound states, the duration time of reaction is naively expected to be significant.
The actual relation between the reaction time and the existence of soliton
will be discussed in the following section (in the context of Fig.~\ref{fig7}).
The high energy cases with $E_s =100.0$~MeV are presented just for the theoretical interest, because the high energy case with its relative velocity 0.81$c$ is more than the application limit of Schr\"odinger type non-relativistic theory.
Roughly speaking, the reliability of non-relativistic framework is believed to be valid for the collisions whose relative velocity is less than 0.50$c$.

\begin{figure}  
  \includegraphics[width=8.0cm]{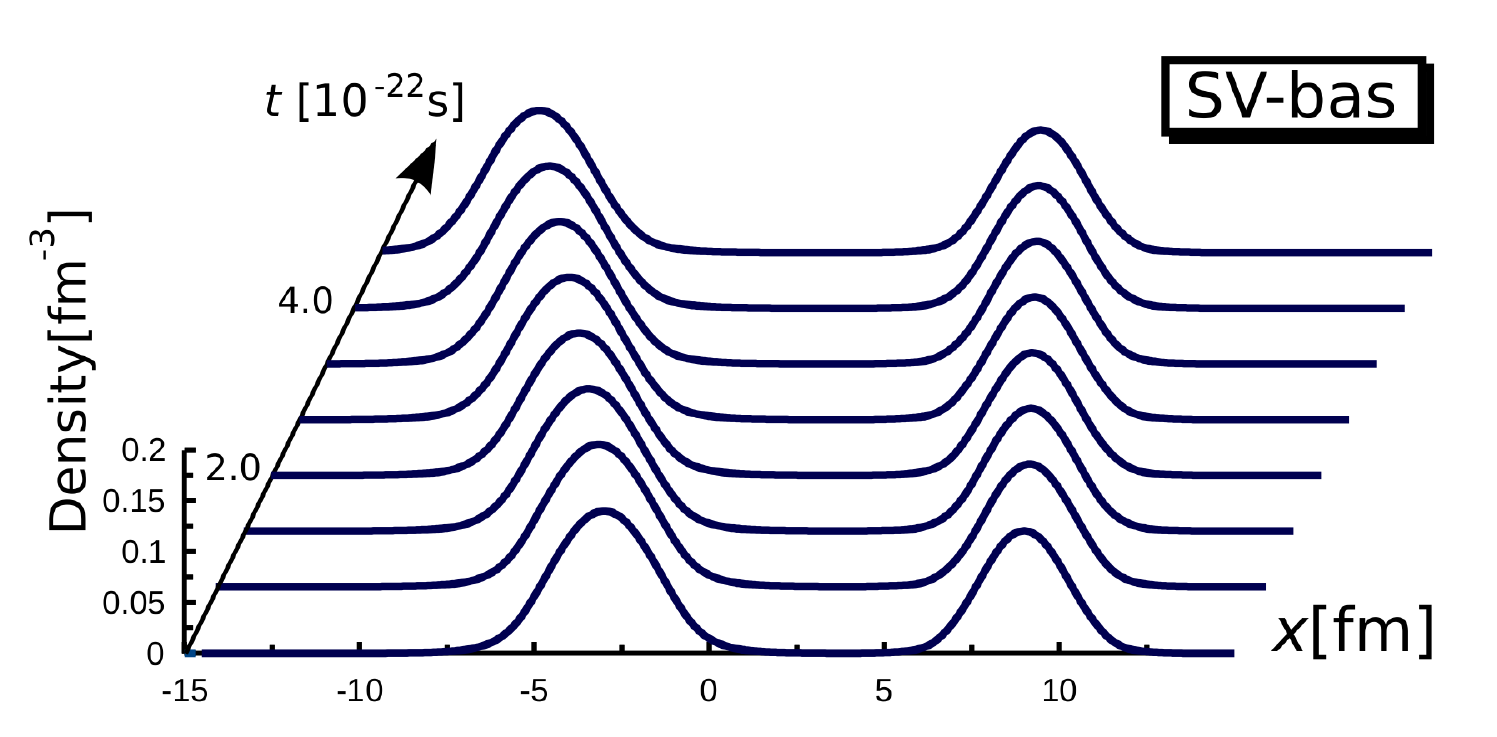}
 \caption{(Color online) Space-time structure of propagating solitary waves. 
At $t = 0.0$~s, bound states of $^8$He and $^4$He are prepared employing SV11111.
$^4$He and $^8$He are placed at (6,0,0) and (-6,0,0), respectively.
In order to avoid the collision between the two ions, the momentum of $^4$He
and $^8$He are especially set to $(-\sqrt{2 M_{4} E_s},0,0)$ and $(-\sqrt{2 M_{8} E_s},0,0)$ respectively, where the reduced incident energy $E_s$ is set to 1.0~MeV.
The time evolution of the density along the axis is shown.
} 
\label{fig2}
\end{figure}

Third, the collision dynamics is calculated for four selected interactions. 
The bound states with the boost are expected to satisfy the property of solitary wave if the obtained bound states are stable.
By adjusting the initial momentum to avoid any collisions between the bound states, the time evolution of bound states with the boost is demonstrated in Fig.~\ref{fig2}.
Figure~\ref{fig2} shows that the bound states are traveling without changing
their local density distribution, if there is no collision.
That is, the bound states with the boost satisfy the property of the solitary waves.
The existence of such decayless propagations of bound states (i.e., the property of solitary wave) is checked for the other three different interactions.
Fig.~\ref{fig2} also shows that the present numerical settings such as space and time incrimination are satisfactory.

\begin{figure*}
\hspace{-13.2cm}
(A) Calculation employing SV11111 (= SV-bas)  \vspace{2.5mm}\\
   \includegraphics[width=17.0cm]{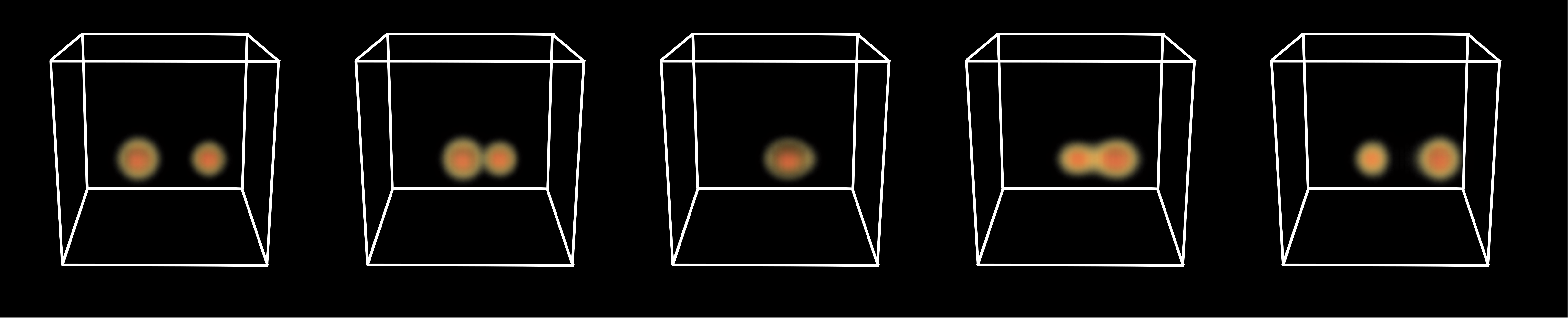}    \vspace{2.5mm}     \\
\hspace{-15cm}
(B) Calculation employing SV11110  \vspace{2.5mm} \\
   \includegraphics[width=17.0cm]{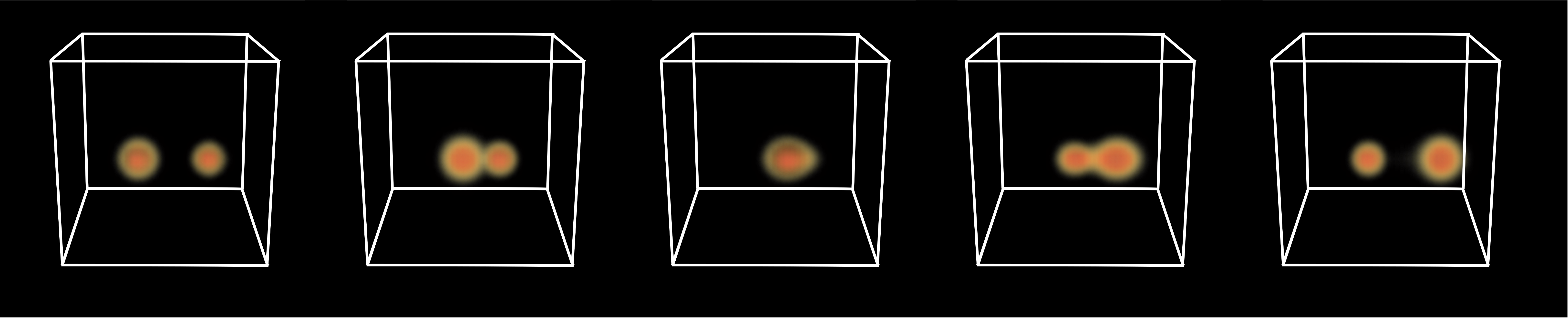}      \\
 \caption{(Color online) Collision dynamics employing SVb11111 interaction
   (Panel (A)) and SV11110 interaction (Panel (B)). 
Dense parts are colored in red, and the dilute parts are colored in yellow.
Snapshots for $t =$ 0, 3.3, 6.6, 9.9, and 13.2 [10$^{-22}$~s] are shown in
each collision.
For both collisions, the reduced incident energy $E_s$ is set to 10.0~MeV, and
the modified impact parameter $b$ is set to 0.0~fm.} 
\label{fig6}
\end{figure*}

Forth, three types of physical quantities are calculated in order to show the existence of the soliton.
To evaluate the transparency rate, we start with the definition of proton and neutron numbers:
\begin{equation}  \begin{array}{ll} 
n_{-1/2,\Omega}(t) = \int_{\Omega} {\displaystyle \sum_j}  |\psi_{-1/2,j}(r,t)|^2 dr^3, \vspace{2mm} \\
n_{+1/2,\Omega}(t) = \int_{\Omega} {\displaystyle \sum_j}  |\psi_{+1/2,j}(r,t)|^2
dr^3,
\end{array}  \end{equation}
and the definition of proton momentum and neutron momentum:
\begin{equation} \begin{array}{ll}
p_{-1/2,\Omega}(t) = \frac{d}{dt} \int_{\Omega}   {\displaystyle \sum_j}
\psi^*_{-1/2,j}(r,t) ~r~ \psi_{-1/2,j}(r,t)  dr^3, \vspace{2mm} \\
p_{+1/2,\Omega}(t) =  \frac{d}{dt} \int_{\Omega} {\displaystyle \sum_j} \psi^*_{+1/2,j}(r,t)
 ~r~ \psi_{+1/2,j}(r,t) dr^3,
\end{array}  \end{equation}  
where $\psi^*_{\pm1/2,j}(r,t)$ is the complex conjugate of
$\psi_{\pm1/2,j}(r,t)$. 
To compare the initial state with final state, the three-dimensional region $\Omega$ is taken as $\Omega_+ = (2,16)
\times (-16,16) \times(-16,16)$ or $\Omega_- = (-16,-2)
\times (-16,16) \times(-16,16)$.
Using these values, the numbers
\[ \begin{array}{ll} 
N_{z,ini} = n_{-1/2,\Omega_-}(0), \vspace{1.5mm}\\
N_{z,fin} = n_{-1/2,\Omega_+}({\bar t}), \vspace{1.5mm}\\
N_{n,ini} = n_{+1/2,\Omega_-}(0), \vspace{1.5mm}\\
N_{n,fin} = n_{+1/2,\Omega_+}({\bar t}), \vspace{1.5mm}\\
P_{z,ini} = p_{-1/2,\Omega_-}(0), \vspace{1.5mm}\\
P_{z,fin} = p_{-1/2,\Omega_+}({\bar t}), \vspace{1.5mm}\\
P_{n,ini} = p_{+1/2,\Omega_-}(0), \vspace{1.5mm}\\
P_{n,fin} = p_{+1/2,\Omega_+}({\bar t}) 
\end{array} \] 
follow,
where ${\bar  t}$ is a certain time being different case by case for the sake
of picking out the transparent components after the collision correctly.
Indices "$ini$" and "$fin$" mean "initial" and "final", respectively.
The values of $N_{z,fin}$, $N_{n,fin}$, $P_{z,fin}$ and $P_{n,fin}$ are assumed to be equal to zero in the presence of fusion.
Note that numbers $N_{z,fin}$ and $N_{n,fin}$ are not generally integers, which is due to the property inherent to the nuclear density functional theory (cf. mean-field property).
Using these quantities the transparency rate of mass is measured by 
\begin{equation} \label{trans1}  \begin{array}{ll} 
T_{\rho,Z} = \frac{N_{z,fin}}{N_{z,ini}}, \quad  
T_{\rho,N} = \frac{N_{n,fin}}{N_{n,ini}}, \quad  
T_{\rho} =\frac{N_{z,fin}+N_{n,fin}}{N_{z,ini}+N_{n,ini}}.
\end{array}  \end{equation}
The two solitary waves are mass-transparent if $T_{\rho,Z}$ and $T_{\rho,N}$
are equal to 1, proton transfer (resp. neutron transfer) from $^4$He to $^8$He
is detected if $T_{\rho,Z}$ (resp. $T_{\rho,N}$) is larger than 1,
and proton transfer (resp. neutron transfer) from $^8$He to $^4$He is detected
if $T_{\rho,Z}$ (resp. $T_{\rho,N}$) is smaller than 1.
Meanwhile the transparency rate of momentum is measured by 
\begin{equation} \label{trans2}  \begin{array}{ll} 
T_{\tau,Z} = \frac{P_{z,fin}}{P_{z,ini}}, \quad 
T_{\tau,N} = \frac{P_{n,fin}}{P_{n,ini}}, \quad 
T_{\tau} = \frac{P_{z,fin}+P_{n,fin}}{P_{z,ini}+P_{n,fin}}.
\end{array}  \end{equation}
Momentum transparency is detected if $T_{\tau}$ is equal
to 1, momentum transfer from $^8$He to $^4$He is detected if it is larger than
1, and that from $^4$He to $^8$He is detected if it is smaller than 1.

Since the nuclei consist of the two components (protons and neutrons), it is interesting to
consider ``when'' and ``how'' charge equilibration appears in association with
the appearance of soliton.
Charge equilibration, which is definitely the mixing process between protons and
neutrons towards the equilibrium of charge, is known to be a quite fast process (see \cite{84freiesleben} for
a review mainly on experiments, and see \cite{10iwata, 12iwata} for the theoretical
aspects).
Among others Iwata {\it et al.}~\cite{10iwata} showed that there is the
upper-limit incident energy for the appearance of the fast charge
equilibration, and the propagation speed of charge equilibration wave is found to be almost constant ($\sim
0.25c$) being independent of the incident energy.
On the other hand the propagation speed of soliton is definitely the same as the
amplitude of the relative velocity.
Charge equilibration rate is measured by
\begin{equation} \label{charge} 
{\mathcal E}_{ce} = \frac {\left| \frac{N_n}{N_z} - \frac{N_{n,ini}}{N_{z,ini}}  \right| -  \left| \frac{N_n}{N_z}  - \frac{N_{n,fin}}{N_{z,fin}}   \right|}{\left| \frac{N_n}{N_z} - \frac{N_{n,ini}}{N_{z,ini}}  \right| },
 \end{equation}
where the neutron-to-proton ratio of the charge equilibrium is denoted by $N_n/N_z = 8/4=2$ and
the initial neutron-to-proton ratio for $^8$He is equal to $N_{ini}/Z_{ini} = 6/2 =3$.
It is readily seen that ${\mathcal E}_{ce} \le 1$ is always true.
Charge equilibrium is achieved if ${\mathcal E}_{ce} = 1$.
Charge equilibration dynamics appears only if $0<{\mathcal E}_{ce} \le 1$.
Charge unequilibration, which corresponds to the dynamics opposite to the charge equilibration, appears if ${\mathcal E}_{ce} < 0$.
Neither charge equilibration nor charge unequilibration appears if ${\mathcal E}_{ce} = 0$. 
Under the appearance of charge equilibration, reaction similar to the
following three typical reactions
are expected to appear,
\begin{equation} \label{cereaction} \begin{array}{ll}
^4{\rm He} ~+~ ^8{\rm He} ~\to~ ^{12}{\rm Be} \\
^4{\rm He} ~+~ ^8{\rm He} ~\to~ ^6{\rm He} ~+~ ^6{\rm He},  \\
^4{\rm He} ~+~ ^8{\rm He} ~\to~  ^9{\rm Li} ~+~ ^3{\rm H},  \\
\end{array} \end{equation}
where the first one is the fusion reaction ($N_{n,fin}/N_{z,fin} = 8/4 =2$), the second one is two neutron
transfer reaction from $^8$He to $^4$He ($N_{n,fin}/N_{z,fin} = 4/2 =2$), and the third one is the one proton
transfer reaction from $^4$He to $^8$He ($N_{n,fin}/N_{z,fin} = 6/3 =2$).

\begin{table}
 \caption{Transparency rate of mass defined by Eq.~\eqref{trans1}.
Each box includes $T_{\rho,Z}$, $T_{\rho,N}$, and $T_{\rho}$ values in this order. }
(A) Central collisions ($b=0.0$~fm)  \vspace{1.5mm}  \\
 \begin{tabular}{||c||c|c|c||} \hline
   Force     & ~$E_s =$1.0~  & $E_s =$10.0~   & $E_s =$100.0  \\   \hline 
   SVb11010 & 1.06,~0.79,~0.85 & 1.00,~0.92,~0.94  & 1.18,~1.05,~1.08  \\   \hline
   SVb11011 & 0.00,~0.00,~0.00 & 1.00,~0.93,~0.94 &  1.24,~1.11,~1.14    \\   \hline
   SVb11110 & 0.00,~0.00,~0.00 & 1.01,~0.93,~0.95  & 1.07,~1.01,~1.02  \\   \hline
   SVb11111 & 0.00,~0.00,~0.00 & 0.95,~0.92,~0.95 &  1.09,~1.01,~1.04  \\   \hline
  \end{tabular}  \vspace{2.5mm} \\
(B) Peripheral collisions ($b=4.0$~fm)  \vspace{1.5mm}   \\
 \begin{tabular}{||c||c|c|c||} \hline
   Force     & ~$E_s =$1.0~  & $E_s =$10.0~   & $E_s =$100.0  \\   \hline 
   SVb11111 & 0.00,~0.00,~0.00 & 1.14,~0.97,~1.02  & 1.18,~1.02,~1.06
   \\   \hline  
 \end{tabular} 
 \label{table2}
\end{table}

\section{Results} \label{sec-res}
We seek soliton solutions to the TDDFT equations Eq~\eqref{nonstationaryaryeq}, where the reduced incident
energy (denoted by $E_s$) and the modified impact parameter (denoted by $b$)
are the two major free parameters for the collision dynamics.

Time evolutions of ``$^4$He + $^8$He'' for $E_s = 1.0$~MeV and $b
=0.0$~fm are demonstrated in Fig.~\ref{fig6}, where SVb11110 and SVb11111
interaction sets are employed.
In both cases two nuclei have a contact around $t= 3.3 \times 10^{-22}$~s, and
separate again around $t= 9.9 \times 10^{-22}$~s.
These two time evolutions entailing a certain level of mass-transparency are
seemingly similar, and the further investigation calculating Eqs.~\eqref{trans1},
\eqref{trans2}, and  \eqref{charge} will see the difference between them.

\begin{figure*}  
 \hspace{0mm} (A) Low energy ($E_s = 1.0$~MeV) \hspace{34mm} (B) Medium energy ($E_s = 10.0$~MeV)   \hspace{40mm} \vspace{1.5mm}  \\
 \includegraphics[width=7.0cm]{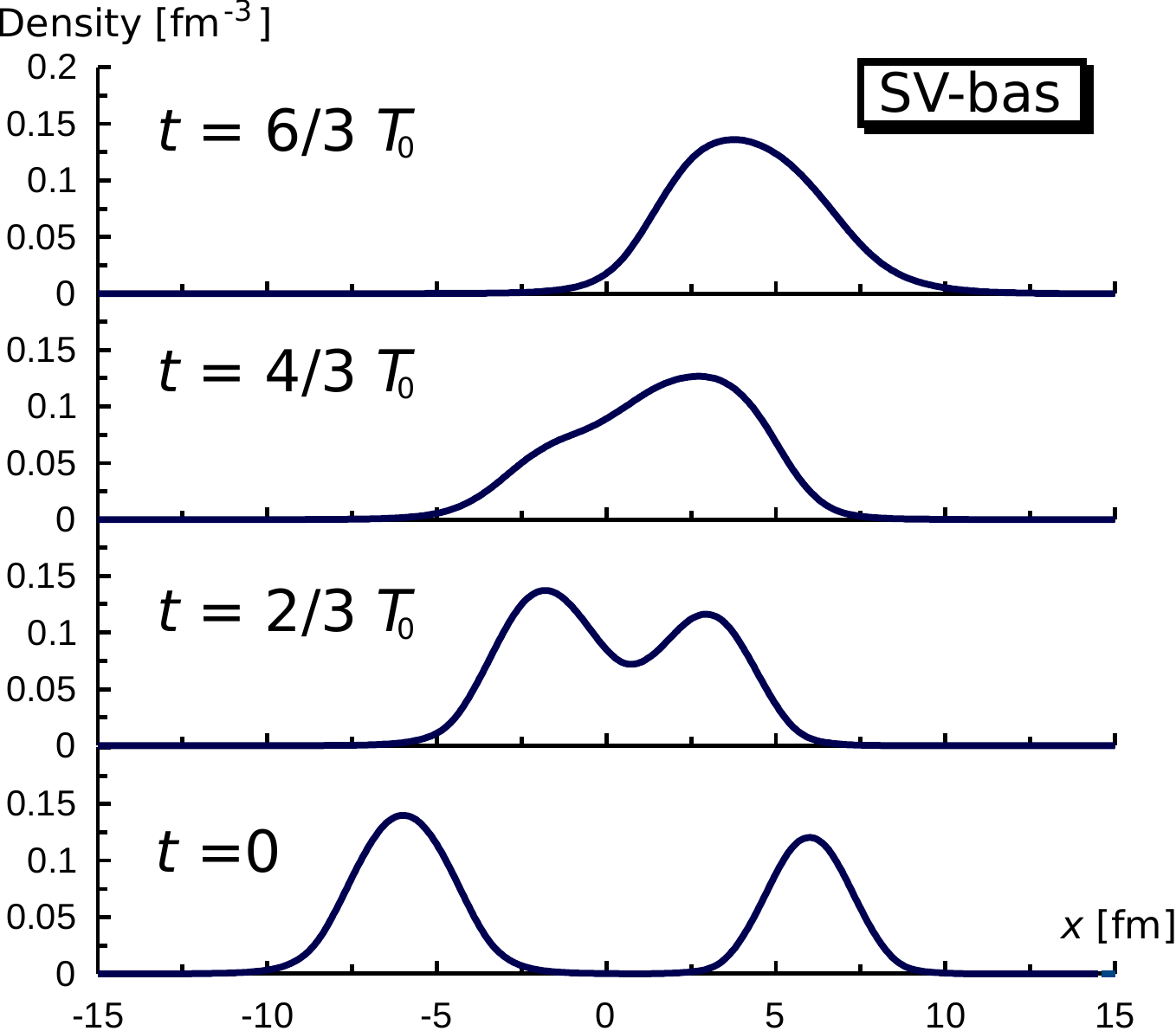} \qquad \qquad
 \includegraphics[width=7.0cm]{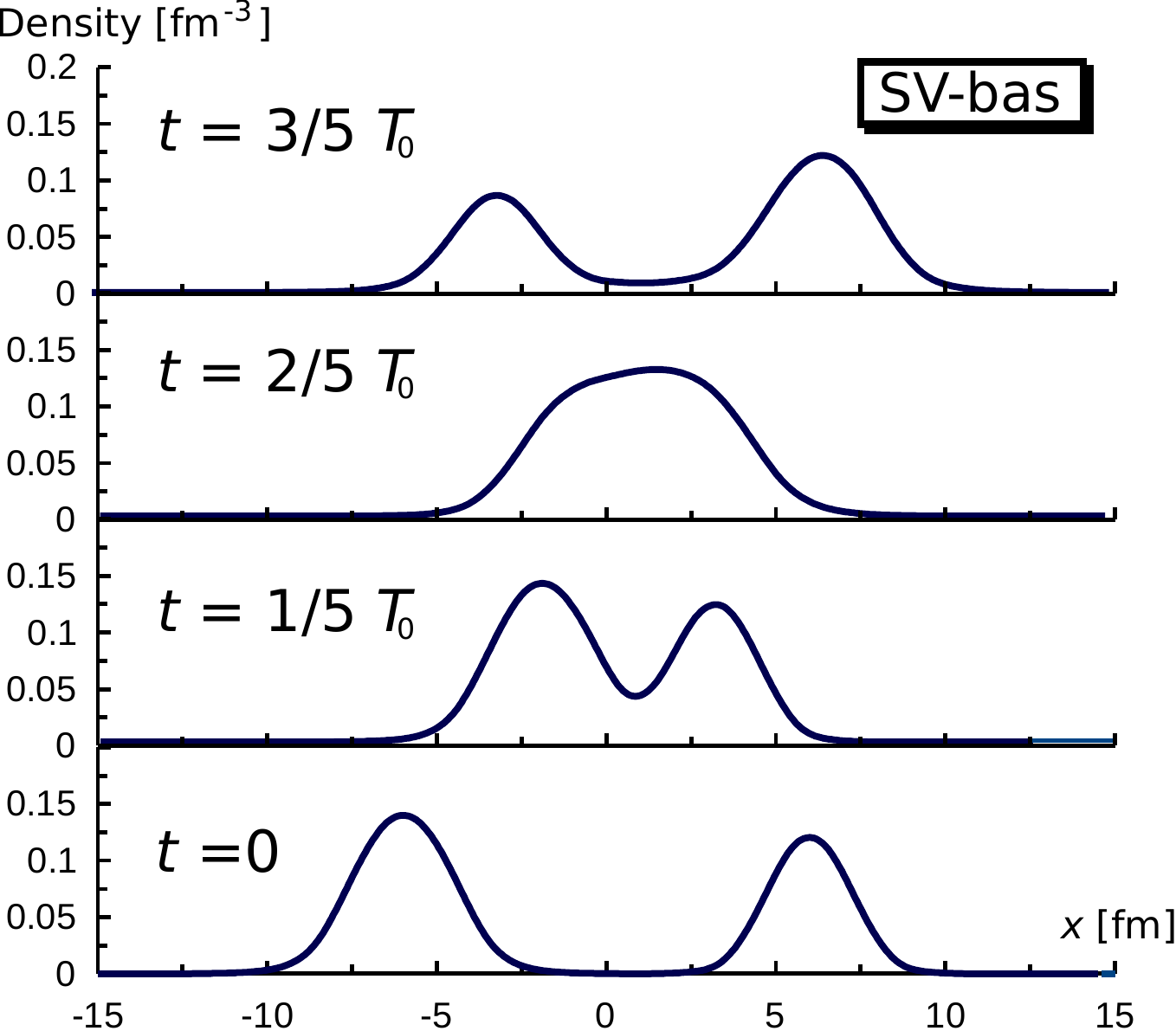} 
\caption{(Color online) Time evolution of ion collisions employing SVb11111 interaction set, where $T_0$ is set to $4.0 \times 10^{-22}$~s.
The time evolution of mass distribution along the collision axis $(x,0,0)$ is depicted for two different incident energies.
Fusion appears in low-energy case (panel (A)), while fragmentation entailing the soliton component appears in higher energy case  (panel (B)). 
} 
\label{fig5}
\end{figure*}

The mass-transparency rate is summarized in Table~\ref{table2}~(A).   
According to the low-energy results ($E_s = 1.0$~MeV), the transparency rate is increased by employing SVb11010 interaction which does not
include the higher order momentum contribution ($t_2$-term) of
the nuclear force and the spin-orbit force ($W_0$-term).
Note again that $t_3$-term has a relatively similar property to $t_0$ term,
because they are independent of the momentum.
According to the medium-energy results ($E_s = 10.0$~MeV), even independent of the
force parameter set, the transparency rate for protons and neutrons are almost
equal to 1.00 and 0.92, respectively.
In these cases the protons are almost transparent as they are.
According to the high energy results ($E_s = 100.0$~MeV), nucleon
transfers from $^4$He to $^8$He are noticed.
Such kind of nucleon transfers enlarge the difference of the masses of the two nuclei.
For peripheral collisions (Table~\ref{table2}~(B)), the transparency rates is thoroughly larger compared to that for the central collisions.
It is seen here that proton transfer from $^4$He to $^8$He is enhanced in peripheral cases.
It is worth noting that peripheral collisions appear more frequently than the central collisions with respect to the contribution to the reaction cross section.

\begin{table}
 \caption{Charge equilibration rate ${\mathcal E}_{ce}$ defined by Eq.~\eqref{charge}.}
(A) Central collisions ($b=0.0$~fm)   \vspace{1.5mm}  \\
 \begin{tabular}{||c||c|c|c||} \hline
   Force     & ~$E_s =$1.0~  & $E_s =$10.0~   & $E_s =$100.0  \\   \hline 
   SVb11010 & 0.76  & 0.23  & 0.33  \\   \hline
   SVb11011 & 1.00  & 0.21  & 0.33   \\   \hline
   SVb11110 & 1.00  & 0.22  & 0.16  \\   \hline
   SVb11111 & 1.00  & 0.27  & 0.22 \\   \hline
  \end{tabular}  \vspace{2.5mm} \\
(B) Peripheral collisions  ($b=4.0$~fm)   \vspace{1.5mm}   \\
 \begin{tabular}{||c||c|c|c||} \hline
  Force     & ~$E_s =$1.0~  & $E_s =$10.0~   & $E_s =$100.0  \\   \hline 
   SVb11111 & 1.00 & 0.43  & 0.41  \\   \hline
  \end{tabular}  \vspace{2.5mm}  \\
  \label{table4}
\end{table}

By making more accurately energy-incremented calculations with incident energies from $E_s=2.0$, 3.0, $\cdots$ to 9.0~MeV, the highest incident energy for the fusion appearance is
located at $E_s = 3.0$~MeV for SVb11011 and SVb11110 interactions, and at $E_s = 4.0$~MeV for SVb11111 interaction.
Fusion is hindered for SVb11011 and SVb11110 interactions compared to SVb11111 interaction.
That is, fusion is enhanced by the highly momentum-dependent term of the nuclear potential and the spin-orbit force.

Charge equilibration rate is summarized in Table~\ref{table4}. 
The charge equilibration appears in all 12 cases shown in Table~\ref{table4},
which is noticed by seeing all the values of ${\mathcal E}_{ce}$ satisfy $0 < {\mathcal E}_{ce} \le 1.00$.
Meanwhile the charge equilibrium is achieved only in three cases with ${\mathcal E}_{ce} = 1.00$.
The high mass-transparency rate of only protons in medium energy cases
(cf. Table~\ref{table2}) is explained by this charge equilibration dynamics, where neutron-transfer
reaction ``$^4{\rm He} ~+~ ^8{\rm He} ~\to~ ^6{\rm He} ~+~ ^6{\rm He}$'' is
preferred among three reactions shown in Eq.~\eqref{cereaction}. 
Depending on the incident energy, charge equilibration dynamics is connected to three different reaction types shown in Eq.~\eqref{cereaction}; the charge equilibration is mostly achieved
by fusion reaction in the low energies, by the neutron transfer from $^8$He to $^4$He in the medium energies, and by the proton transfer from $^4$He to $^8$He in the high energies.
Such an energy-dependent trend is also supported by the prediction made by the classification of charge equilibration dynamics~\cite{10iwata-np}.
By comparing the results for the medium energy ($E_s=10.0$~MeV) with those for the
high energy ($E_s=100.0$~MeV), the charge equilibration rate becomes smaller
for the higher energy if SVb11110 or SVb11111 interaction is employed, but it
becomes larger for the higher energy if SVb11010 or SVb11011 interaction is employed.
It is actually due to the highly momentum-dependent term of the
nuclear potential ($t_2$-term).
Charge equilibration is suppressed by the highly momentum-dependent term for higher energies, even though highly momentum-dependent terms are expected to increase the interaction between the two solitary waves for higher energies.
By comparing the results by SVb11111 with those by SVb11110, the spin-orbit
force works as increasing the charge equilibration rate.
According to Table~\ref{table4}~(B), it is clearly seen that the charge
equilibration is enhanced for peripheral collisions.

In addition to the mass-transparency, the momentum-transparency is another
degree of freedom to identify the existence of soliton. 
Momentum-transparency rate is summarized in Table~\ref{table3}.
Momentum-transparency rates are
calculated to be close to 1.0 (e.g., $0.80 \le T_{\tau} \le 1.20$) only in the medium energy cases ($E_s = 10.0$~MeV).
It shows that the momentum-transparency is not necessarily close to 1.0, even though the mass
transparency is almost close to 1.0.
That is, with respect to the synthesis of chemical elements, the
momentum transfer contributes to suppress the appearance of soliton more than the
mass transfer.

The shape is not measured by any transparency rates introduced in Eqs.~\eqref{trans1}, \eqref{trans2}, \eqref{charge}.
Time evolution of the shapes is shown in Fig.~\ref{fig5}.
Fusion and fragmentation appear depending on the incident energy.
The chemical element  "beryllium" (more precisely $^{12}$Be) is produced in fusion reaction (Panel (A)) in which no soliton property remains.
The shapes of localized density concentration at $t=3/5 T_0$ of Panel (B) are not so different from those in the initial state ($t= 0$). 
Panel (B) shows that neutrons and protons self-organize into a bound state with a certain excitation energy even after the collision, once a connected mass distribution with certain proton and neutron numbers are given.

\begin{table}
 \caption{Transparency rate of momentum $T_{\tau}$ defined by Eq.~\eqref{trans2}.
\vspace{1.5mm}}
 \begin{tabular}{||c||c|c|c||} \hline
   Force     & ~$E_s =$1.0~  & $E_s =$10.0~   & $E_s =$100.0  \\   \hline 
   SVb11010 & 0.42  & 0.87  & 1.75  \\   \hline
   SVb11011 & 0.00 & 0.88 & 1.91  \\   \hline
   SVb11110 & 0.00  & 0.90  & 1.58  \\   \hline
   SVb11111 & 0.00   & 0.85 & 1.53 \\   \hline
  \end{tabular}  \vspace{2.5mm} \\
 \label{table3}
\end{table}

\begin{figure}   
\caption{(Color online) Energy-dependence of the total transparency rate $T$,
  the charge equilibration rate ${\mathcal E}_{ce}$ and the reaction time is shown for SVb11111 interaction, where the total transparency rate at $E_s = 1$~MeV is equal to 0.0.
The reaction time, which means the duration time interval of the reaction, is estimated by Eq.~\eqref{timestim}. 
The horizontal axis is plotted using the logarithmic scale.
The left and right vertical axes are for the total transparency rate and the
charge equilibration rate, respectively.
There is no axis for the reaction time, and values are shown besides the
curve on three points, instead.     
  } 
\hspace{-0cm}
 \includegraphics[width=9.50cm]{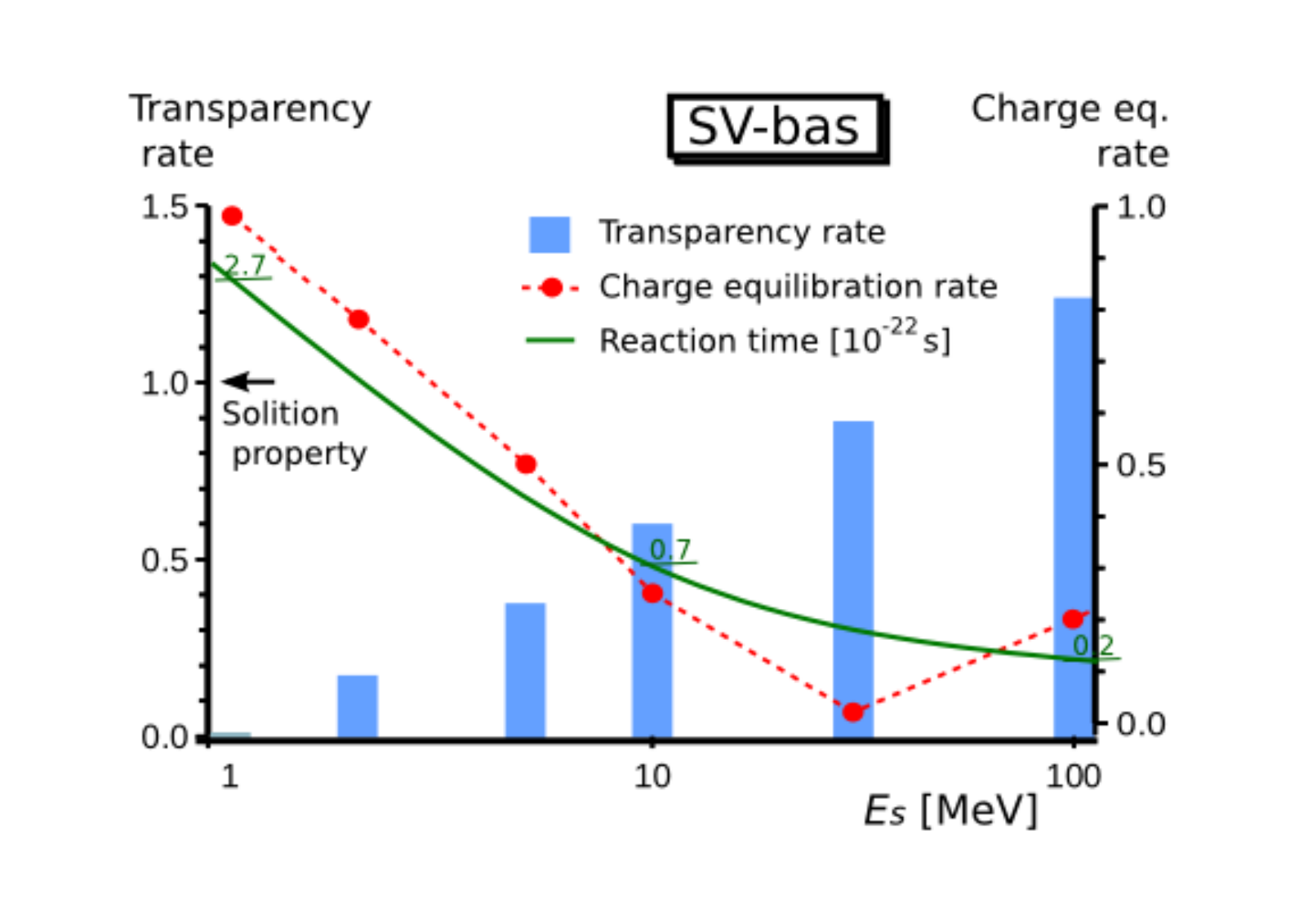} 
\label{fig7}
\end{figure}

As the amount of momentum-transparency has been claimed to be so small in ion reactions \cite{82galin,83raha}, one of the purpose of this paper is to find out the reason why the momentum transparency is small.
The total transparency rate including the
mass-transparency and the momentum-transparency is defined by
\begin{equation}
 {\mathcal T} = T_{\rho,Z} \times T_{\rho,N} \times T_{\tau,Z} \times T_{\tau,N}.
\end{equation} 
The total transparency is detected if ${\mathcal T}$ is equal to 1.
As a criterion in this paper, rather pure soliton existence is suggested if $0.80 \le {\mathcal T} \le 1.20$, where the amplitude of the contained amount of soliton component is estimated by the value $|1-{\mathcal T}|$.
This criterion is only a necessary condition for the soliton existence,
because it does not say nothing about the shape.
Energy-dependence of the total transparency rate, the charge equilibration
rate and the reaction time are compared in Fig.~\ref{fig7}.
First, the significant charge equilibration is notice only for the lower energy cases ($E_s \le 10$~MeV).
Second, the soliton is suggested to exist in relatively high energy collision
(around $E_s$=30~MeV) at which the charge equilibration rate has a minimum
value with respect to the energy.
Such a correspondence implies that the
appearance of charge equilibration is the primary factor of the soliton suppression.
Third, more charge equilibration does not necessarily result in the
suppression of soliton (see very high
mass-transparency of protons at $E_s = 10.0$~MeV shown in Table~\ref{table2}
even under the appearance of charge equilibration).
Forth, less reaction time cannot
be the ultimate factor of the soliton appearance (see
$E_s=100$~MeV case) even though much reaction time is naively expected to bring about more interaction between
the solitary waves.
Indeed, the value of $|1-{\mathcal T}|$ is almost equal to 0.25 for the reaction time 0.2$\times 10^{-22}$~s ($E_s = 100.0$~MeV), while $|1-{\mathcal T}|$ is roughly equal to $0.0$ for the reaction time 0.35$\times 10^{-22}$~s ($E_s = 10.0$~MeV).
Fifth, we found out that the energy-dependence of the total transparency has a clear logarithmic dependence.

\section{Conclusion}
Based on a systematic three-dimensional time-dependent density functional
calculation, the existence of soliton has been studied in ion collisions yielding several kinds of light chemical elements. 
The contained amount of soliton component has been confirmed to be energy-dependent.
In this sense nuclei can be solitons conditionally.
Indeed, the existence energy region of soliton is suggested to be located
around the several 10~MeV per nucleon in the center-of-mass frame, where the reduced incident energy $E_s$ gives an approximation to the center-of-mass energy per nucleon of the total colliding system.
As a result the existence of nuclear transparent incident energy (nuclear transparent temperature), at which the reaction cross section is almost equal to zero, is proposed.
This is nothing but a nonlinear effect, which must be important to understand the structure of the celestial bodies such as stars and neutron stars.

The fast wave realizing the fast charge equilibration is  the primary factor of the soliton suppression; the soliton is hindered if the incident energy is less than the upper-limit energy of the fast charge equilibration ($E_s \sim 10$MeV in this case~\cite{10iwata}).
However charge equilibration does not necessarily suppress the soliton wave.
Indeed, protons have perfect mass-transparency ($T_{\rho,Z} \sim 1.0$) in case of $E_s = 10$~MeV (see cases of SVb11010 and SVb11011 interactions in Table~\ref{table2}), which cannot be
explained without taking into account the charge equilibration dynamics exchanging only neutrons.
As a result a new concept of the cooperative co-existence of soliton propagation and charge equilibration is proposed.

A motivation of this paper is, with respect to the synthesis of chemical elements, to identify the origin of soliton suppression in the components of the effective nuclear force.
First of all,  less reaction time of the two colliding solitary waves does not necessarily lead to the enhancement of soliton existence. 
The larger force-dependence of the soliton existence can be found in the momentum-transparency than in
the mass-transparency.
In this context the spin-orbit force, whose origin is the special relativity effect, plays a crucial role (compare SVb11111
and SVb11110 in case of $E_s = 10$~MeV shown in Table~\ref{table3}).
Rather pure soliton wave is generated by the momentum independent part of
the nuclear force (compare SVb11111 and SVb11011 in case of $E_s = 10$~MeV
shown in Table~\ref{table3}), but it is contaminated by the momentum
dependent parts including the spin-orbit force.
Note that the spin-orbit force was shown to increase the charge equilibration (rate) by increasing the spin polarization~\cite{11iwata}.
As a result the momentum exchange has been confirmed to be crucial to the suppression of soliton.

Finally it is worth while to refer to two effects, which are not taken into account in this paper.
In the higher energies the collision between
nucleon, which gives rise to the effect similar to the friction, is expected
to appear and to break the unitary time evolution.
Nucleon-nucleon collisions are negligible for ion collisions with lower
energies ($E_s \le 10$~MeV) at least, because mean free path of nucleons
in the nuclear matter is calculated to be larger than 20~fm if the energy is less than the Fermi energy~\cite{80collins} (the corresponding $E_s$  is roughly equal to be 10~MeV based on the calculation method shown in \cite{13iwata}).
Another effect is the special relativity effect, which is effectively taken
into account in the force parameterization including the spin-orbit force even
in our non-relativistic framework, but it is believed to be more correctly
treated when the amplitude of the relative velocity of the collision is larger than $0.50c$ (for the theoretical study dealing with ion reactions with high energies, see for example \cite{02danielewicz}).
Charge equilibration rate and the high total transparency rate at
$E_s = 100$~MeV in Fig.~\ref{fig7} should be affected by these two effects.

This work was supported by HPCI Strategic Programs for Innovative Research (SPIRE)
Field 5 ``The origin of matter and the universe", and by the Helmholz alliance HA216/EMMI.
Numerical computation was carried out at the Yukawa Institute Computer Facility (Kyoto University).
The author is grateful to Prof. Naoyuki Itagaki for some comments.

\end{document}